\documentclass[12pt]{article}
\usepackage{amsmath,amsthm}
\usepackage{amssymb}

\usepackage{graphicx}

\newcommand{\be}{\begin{align}}
\newcommand{\ee}{\end{align}}
\newcommand{\ben}{\begin{align*}}
\newcommand{\een}{\end{align*}}

\usepackage{url}
\usepackage{slashbox}
\usepackage[makeroom]{cancel}

\def\eps{\varepsilon}

\begin{document}

\title{Statistical mechanics approach in the counting of integer partitions}

\author{\textit{Andrij Rovenchak}\protect\\
Department for Theoretical Physics,\protect\\
Ivan Franko National University of Lviv,\protect\\
12, Drahomanov St., Lviv, UA-79005, Ukraine\protect\\
e-mail: andrij.rovenchak@gmail.com
}

\maketitle

\begin{abstract}
The treatment of the number-theoretical problem of integer partitions within the approach of statistical mechanics is discussed. Historical overview is given and known asymptotic results for linear and plane partitions are reproduced. From numerical analysis of restricted plane partitions an asymptotic formula is conjectured for an intermediate number of parts.


\end{abstract}

\section{Introduction} 
The problem of integer partitions has quite a long story. For the first time it appeared, most evidently, in a letter of Leibniz to Jacob Bernoulli from 1674 \cite{Leibniz:1674}. Several decades later, Leonhard Euler significantly contributed to study of this problem \cite{Euler:1751,Euler:1753}, having been initially drawn into the problem by Philipp Naud\'e.

A partition of a positive integer $n$ is a way of writing $n$ as a sum of positive integers (called \emph{parts}), where the order of the summands is not significant \cite{Andrews:1976}, possibly subject to one or more additional constraints. The number of partitions is denoted $p(n)$ and is called \emph{partition function} in number theory, but we will preserve this term to the \emph{Zustandssumme} in statistical physics.

Consider partitions for first several integers. For the sake of convenience and compatibility with further definitions, the parts are written in a non-increasing order:
\ben
1\ &:\quad 1,\\[-1pt]
2\ &:\quad 2 = 1+1,\\[-1pt]
3\ &:\quad 3 = 2+1 = 1+1+1,\\[-1pt]
4\ &:\quad 4 = 3+1 = 2+2 = 2+1+1 = 1+1+1+1,\\[-1pt]
5\ &:\quad 5 = 4+1 = 3+2 = 3+1+1 = 2+2+1 = 2+1+1+1 = 1+1+1+1+1.
\end{align*}
So, $p(1)=1$, $p(2) = 2$, $p(3)=3$, $p(4) = 5$, $p(5)=7$. In fact, $p(n)$ is a rapidly growing function: $p(10)=42$, $p(100) = 190\,569\,292$, and $p(200)=3\,972\,999\,029\,388$ \cite[p.~836]{Abramowitz&Stegun:1972}.

The partitions described above can be called \emph{simple, one-dimensional,} or \emph{linear}. In 1917, Hardy and Ramanujan \cite{Hardy&Ramanujan:1917,Hardy&Ramanujan:1918} reported an asymptotic estimation for the number of integer partitions at large $n$ in the form:
\be\label{eq-HR}
p(n) = {1 \over 4\sqrt3\, n}\exp\left\{\pi\sqrt{2n\over3}\right\}, 
\end{align}
Hardy and Ramanujan also obtained an asymptotic expansion for the $p(n)$ function. Two decades later, Rademacher provided a convergent series for $p(n)$ \cite{Rademacher:1938}.

The so-called \emph{two-dimensional} or \emph{plane partition} of a positive integer number $n$ is a two-dimensional array of nonnegative integers $n_{ij}$ subject to a nonincreasing condition across rows and columns, such that \cite{Andrews:1976}
\be
n = \sum_{i,j> 0} n_{ij},\qquad
\textrm{where}\quad 
n_{i_1j_1}\geq n_{i_2j_2}\quad
\textrm{whenever}\quad i_1\leq i_2, j_1 \leq j_2.
\end{align}

For instance, the two-dimensional partitions of 3 are:
$$
3,\ \ \ 2\ 1,\ \ \ {2\atop1},\ \ \ 1\ 1\ 1,\ \ \ {1\ 1\atop 1\ \ }, \ \ \ 
\begin{array}{c}
1\\ 1\\ 1
\end{array}
$$
and the two-dimensional partitions of 4 are:
\vspace*{-0.1cm}
$$
4,\ \ \ 3\ 1,\ \ \ {3\atop1},\ \ \ 2\ 2,\ \ \ {2\atop2},
\ \ \ 2\ 1\ 1,\ \ \ {2\ 1\atop 1\ \ }, \ \ \ 
\begin{array}{c}
2\\ 1\\ 1
\end{array},\ \ \ 
1\ 1\ 1\ 1, \ \ \ 
\begin{array}{l}
1\ 1\ 1\\ 1
\end{array},\ \ \ 
{1\ 1\atop1\ 1},\ \ \ 
\begin{array}{l}
1\ 1\\ 1\\ 1
\end{array},\ \ \ 
\begin{array}{c}
1\\ 1\\ 1\\ 1
\end{array}.
$$
\vspace*{-0.1cm}%
The number of different plane partitions of $n$ is further denoted as $p^{\rm 2D}(n)$; in the above examples $p^{\rm 2D}(3)=6$ and $p^{\rm 2D}(4)=13$. 

Significant contributions to the study of plane partitions were made by  MacMahon \cite{MacMahon:1896,MacMahon:1915}. The estimation for the asymptotic behavior of the number of plane partitions was given in 1931 by Wright as follows \cite{Wright:1931}:
\be\label{p2d-Wright}
p^{\rm 2D}(n)={[{2\zeta(3)}]^{7/36} \over  \sqrt {6\pi} }
 n^{-25/36}
\exp\left\{
{3\over 2} \left[2\zeta(3)\right]^{1/3} n^{2/3} + \zeta'(-1) \right\}.
\end{align}

Integer partitions discussed so far are called \emph{unrestricted}. However, parts of a partition can be subject to various constraints: one may require them to be odd/even, limit the range of values, etc. The variety of constraints for plane partitions is even richer and can concern their shape with respect to rows and columns. The simplest condition is to limit the number of parts, and such \emph{restricted partitions} with the number of parts not exceeding $N$ will be denoted further as $p_N(n)$ for linear and $p^{2\rm D}_N(n)$ for plane partitions.

Asymptotic estimation for the number of the restricted linear partitions was first obtained by Erd\H{o}s and Lehner \cite{Erdos&Lehner:1941} as follows:
\be\label{eq:EL-ln}
\ln\frac{p_N(n)}{p(n)}=-\frac{\sqrt{6n}}{\pi}\,e^{-\pi N/\sqrt{6n}},
\end{align}
A similar result for restricted plane partitions seems not to find a proper reflection in the literature. The relevant asymptotic formulas obtained by the present author \cite{Rovenchak:2014TMF} are discussed in Sec.~\ref{sec:Restricted}.

\section{Physical analogy and method description}\label{sec:Phys}

The connection between a physical problem and integer partitions was first noted by Bohr and Kalckar in 1937 \cite{Bohr&Kalckar:1937} with respect to the calculation of the density of energy levels in heavy nuclei. In the same year, Van~Lier and Uhlenbeck noted the links between counting microstates of the systems obeying Bose or Fermi statistics and some integer partition problems \cite{VanLier&Uhlenbeck:1937}.
Quite comprehensive bibliography on the relation between physical problems and integer partitions can be found at {\normalsize\url{http://empslocal.ex.ac.uk/people/staff/mrwatkin//zeta/partitioning.htm}}.
For a short review, see also \cite{Debnath:1987}.

Studies of integer partitions utilizing the approach of statistical mechanics with re\-spect to ensembles of quantum particles were made by Auluck and Kothari \cite{Auluck&Kothari:1946}, Temperley \cite{Temperley:1949}, Nanda \cite{Nanda:1951}, Dutta \cite{Dutta:1953,Dutta:1956}, and many others. In recent decades, the fluctuations in quantum ensembles and finite-size effects were in the focus \cite{Chatterjee&Diaconis:2014,Roccia&Leboeuf:2010,Rovenchak:2009FNT,Tran_etal:2004}, see also a series of papers by Grossmann and Holthaus \cite{Grossmann&Holthaus:1996,Grossmann&Holthaus:1997,Grossmann&Holthaus:1999}.
Plane partitions are linked to problems in theory of crystals, random walks on lattices, various problems of statistical physics, etc. \cite{Bogoliubov:2007}.

The approach presented below is mostly based on \cite{Tran_etal:2004}.
The underlying idea of the analogy between counting the number of partitions and number of microstates of a quantum system of bosonic harmonic oscillators is quite transparent. Consider for simplicity the first nontrivial example of partitions of $n=3$:
$$3,\qquad 2+1,\qquad 1+1+1.$$
So, as it was mentioned in the Introduction, $p(3)=3$.

\begin{figure}[h]
\centerline{\includegraphics[scale=1.0]{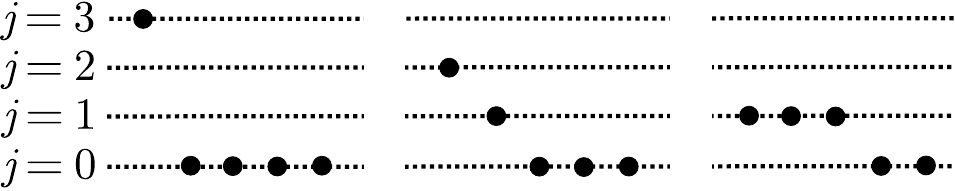}}
\caption{All possible realization of the state with energy $E=3\hbar\omega$ in the system of harmonic oscillators.
}\label{fig:count}
\end{figure}

Consider then the system of one-dimensional quantum harmonic oscillators with frequency $\omega$ obeying the Bose statistics. Let us count the number of ways to distribute the energy of $E=3\hbar\omega$ in this system. Each oscillator can occupy energy levels given by $\eps_j = \hbar\omega (j+1/2)$, where the ground-state energy can be dropped off yielding
\be
\eps_j = \hbar\omega j, \qquad j=0,1,2,3\ldots\,.
\end{align}
The possible ways to achieve the energy $E=3\hbar\omega$ are:
\begin{itemize}
\item to put one particle into excited state with $j=3$;
\item to put one particle to the state with $j=2$ and another one into the $j=1$ state. Note that quantum particles are indistinguishable!
\item to put any three particles into the state $j=1$ each.
\end{itemize}
All the remaining particles remain in the ground state $j=0$. The possibility for several particles to occupy a single energy level is ensured by the Bose statistics. For illustration, see Fig.~\ref{fig:count}.

Therefore, the property that the order of summands is not significant in partitions is equivalent to the indistinguishability of quantum particles. The number of identical parts can reach any value as with the occupation of a quantum state in the Bose statistics. Moreover, it is obvious that unrestricted partitions correspond to a system with an infinite number of bosons and restricted partitions (meaning, with a restricted number of parts) correspond to a system with a finite number of bosons.

Some other types of partitions also can be related to physical models. For instance, the so called \emph{distinct partitions}, where all the parts are different, correspond to fermionic systems.  An example of such a partition is given by
\ben
5\ :\quad 5 = 4+1 = 3+2.
\end{align*}

If all the parts are less or equal to some $s$, such partitions can be modeled by a system obeying the Gentile statistics \cite{Rovenchak:2009FNT}. An example of such a partition is shown below:
\ben
5\ &:\quad  1+1+1+1+1
  = 2+1+1+1\\
  &\quad= 3+1+1\ 
  \cancel{= 4+1}
  = 2+2+1
  = 3+2\ 
   \cancel{= 5}
\qquad\textrm{for}\ s=3.
\end{align*}

One can also consider partitions into powers of integers, e.g.: 
$$5\ :\quad 1^2+1^2+1^2+1^2+1^2 = 2^2+1^2,$$
which correspond to a system of free bosons due to the free particle spectrum $\eps_j\propto j^2$. Other generalization in the problem of partitions are also possible.

The method to link the problem in statistical mechanics to its counterpart in number theory is as follows. The partition function (Zustandssumme) $Z(\beta)$ is connected to the number of microstates ${\it\Gamma}(E)$ by the Laplace transform \cite{Comtet_etal:2007,Grossmann&Holthaus:1997,Tran_etal:2004}:
\be \label{ZLaplace} Z(\beta) = \int_0^\infty {\it\Gamma}(E)\,
e^{-\beta E}\, dE,
\qquad
{\it\Gamma}(E) =
{1\over2\pi i} \int_{-i\infty}^{+i\infty} Z(\beta)\, e^{\beta E}\,
d\beta,
\end{align}
where $\beta = 1/T$ is the inverse temperature.

After some transformations, using the method of steepest descent, the number of microstates is obtained as 
\be\label{eq:GammaE}
{\it\Gamma}(E) = \frac{{\rm e}^{S(\beta_0)}}{\sqrt{2\pi S''(\beta_0)}},
\end{align}
where the entropy is given by
\be
S(\beta) = \beta E + \ln Z(\beta)
\end{align}
and $\beta_0$ is the stationary point, so that
\be
S'(\beta_0) = 0.
\end{align}
 
From this expression, estimations for the number of different types of integer partitions can be obtained by considering partition functions of respective physical systems and subsequently substituting the energy $E$ with an integer $n$:
\be
E\to n,\qquad {\it\Gamma}(E)\to p(n).
\end{align}
In the following sections, the technique described above is applied to linear and plane partitions, both unrestricted and restricted.

\section{Application to unrestricted partitions}
\subsection{One-dimensional case} For this problem we expect the asymptotic result of Hardy and Ramanujan. 

In the case of one-dimensional harmonic oscillators obeying the Bose statistics the partition function of a system with an infinite number of particles reads
\be
Z(\beta) = \prod_{j=1}^\infty \frac{1}{1-e^{-\beta\hbar\omega j}},
\end{align}
For simplicity, let further the unit of energy and temperature $\hbar\omega=1$.

The entropy of such a system is thus
\be
S(\beta) = \beta E + \ln Z(\beta) = \beta E - 
\sum_{j=1}^\infty \ln\left(1-e^{-\beta j}\right).
\end{align}

\def\ds{\displaystyle}

Using the Euler--Maclaurin formula to evaluate the sum, cf. \cite[p.~16]{Abramowitz&Stegun:1972},
\be
\sum_{j=1}^\infty f(j)=\int_1^\infty f(x)\,dx + \frac{f(1)}{2}-
\frac{1}{12}\,f'(1)+\ldots,
\end{align}
one obtains in the limit of $\beta\to0$ (which is relevant to the problem as we expect an asymptotic expression for large values of energy $E$):
\be
S(\beta) = \frac{\pi^2}{6\beta} +
\frac{1}{2}\ln\beta - \frac{1}{2}\ln2\pi + 
\left(E-\frac{1}{24}\right)\beta + \ldots\,.
\end{align}
Note the appearance of the 1/24 fraction near energy: such a subtle correction is present in the formula for the number of partitions beyond the main asymptotic \cite{Hardy&Ramanujan:1918,Rademacher:1938}. In the limit of large $E$ or equivalently -- large $n$ -- it can be dropped.

Stationary point satisfying the condition $S'(\beta_0)=0$ is
\be
\beta_0 = \frac{\pi}{\sqrt{6E}}.
\end{align}
So, with $\ds S''(\beta_0) = \frac{2\sqrt6}{\pi}E^{3/2}$ we obtain:
\be
{\it\Gamma}(E) = \frac{1}{4\sqrt3 \,E}e^{\pi\sqrt{2E/3}}
\qquad\textrm{yielding}\qquad
p(n) = \frac{1}{4\sqrt3 \,n}e^{\pi\sqrt{2n/3}},
\end{align}
which is precisely the result of Hardy and Ramanujan.

\subsection{Two-dimensional case} For this problem, Wright's result (\ref{p2d-Wright}) can be expected.

The derivation of an asymptotic expression is similar to the one-dimensional case with minor modifications. A system of two-dimensional  isotropic harmonic oscillators should be considered. Taking into account the $j$-fold degeneracy of the $j$th energy level, we have for entropy
\be
S(\beta) &= \beta E + \ln Z(\beta) = \beta E - 
\sum_{j=1}^\infty j\ln\left(1-e^{-\beta j}\right) \nonumber\\
&= \beta E +\frac{\zeta(3)}{\beta^2}+\frac{1}{12}\ln\beta-\frac16
+\ldots\,.
\end{align}

Stationary point in this case is
\be
\beta_0 = \left(\frac{2\zeta(3)}{E}\right)^{1/3},
\end{align}
where $\zeta(x)$ is the Riemann zeta-function.

With $\ds S''(\beta_0) = \frac{3}{[2\zeta(3)]^{1/3}}E^{4/3}$ we obtain the number of microstates:
\be
{\it\Gamma}^{\rm2D}(E) = 
\frac{[2\zeta(3)]^{7/36}}{\sqrt{6\pi}}E^{-25/36}
\exp\left\{\frac32[2\zeta(3)]^{1/3}E^{2/3}-\frac16\right\}
\end{align}
So,
\be
p^{\rm2D}(n) = 
\frac{[2\zeta(3)]^{7/36}}{\sqrt{6\pi}}n^{-25/36}
\exp\left\{\frac32[2\zeta(3)]^{1/3}n^{2/3}-\frac16\right\},
\end{align}
which is slightly different from Wright's result (\ref{p2d-Wright}):
\be\label{p2d-Wright'}
p^{\rm2D}_{\rm Wright}(n) = 
\frac{[2\zeta(3)]^{7/36}}{\sqrt{6\pi}}n^{-25/36}
\exp\left\{\frac32[2\zeta(3)]^{1/3}n^{2/3}+c\right\},
\end{align}
where
\be
c = \zeta'(-1) = -0.165421\ldots \quad\textrm{versus}\quad
-\frac16 = -0.166666\ldots
\end{align}

To recover the correct value of $c$, the asymptotic series in the Euler--Maclaurin formula \cite{Apostol:1999} must be summed:
\be
\sum_{j=1}^\infty f(j)=\int_1^\infty f(x)\,dx + \frac{f(1)}{2}-
\sum_{k=1}^{k_{\max}} \frac{B_{2k}}{(2k)!}\,f^{(2k-1)}(1)+R,
\end{align}
where $B_{2k}$ are the Bernoulli numbers and $R$ is the remainder term. For convenience, terms corresponding to the values of the function $f(x)$ and its derivative at infinity were dropped as they are zero in the case under consideration.

\begin{table}[h]
\caption{Evaluation of terms in the asymptotic series for two-dimensional partitions}\label{tab:1}\medskip
\begin{center}
\begin{tabular}{cr}
\hline
$k$ & term value \\
\hline
2 & $-0.0013889$\\
3 &$+0.0001984$\\
\textbf{4} & $-0.0000992$\\
5 &$+0.0001052$\\
6 & $-0.0001918$\\
\hline
\end{tabular}
\end{center}
\end{table}

To sum up an asymptotic series with terms having alternating signs it is sufficient to stop at the term with the least absolute value ($k=4$ in Table~\ref{tab:1}). Thus, with up to $B_6$ and $B_8$ terms, the following values of the $c$ constant are obtained instead of $-1/6$:
\ben
\begin{array}{r}
\ds-\frac{139}{840}=-0.165476\ldots\\[16pt]
\ds-\frac{1667}{10\,080}=-0.165377\ldots
\end{array}\quad\textrm{versus}\quad c = \zeta'(-1) = -0.165421\ldots
\,.
\end{align*}

In fact, the exact value of $c=\zeta'(-1)$ can be obtained within the model of two-dimensional harmonic oscillators but a much more sophisticated approach should be utilized as shown by Nanda \cite{Nanda:1951}.

\subsection{Correction to the main asymptotic}

Adding one more term in the expansion of entropy
\be
S(\beta) \simeq 
S(\beta_0) + \frac{1}{2!}S''(\beta_0)(\beta-\beta_0)^2
+ \frac{1}{3!}S'''(\beta_0)(\beta-\beta_0)^3,
\end{align}
yields the number of microstates evaluated by the method of steepest descend as follows \cite{Prokhorov&Rovenchak:2012}:
\be
{\it \Gamma}(E) = \frac{e^{S(\beta_0)}}{2\pi}\,
\frac{2S''(\beta_0)}{\sqrt3\,|S'''(\beta_0)|}
\exp\left\{ \frac{[S''(\beta_0)]^3}{3[S'''(\beta_0)]^2} \right\}
K_{1/3}\left( \frac{[S''(\beta_0)]^3}{3[S'''(\beta_0)]^2} \right),
\end{align}
where $K_\nu(x)$ is the Macdonald function (modified Bessel function of the second kind).

\begin{figure}[h]
\centerline{\includegraphics[scale=.4,angle=-90]{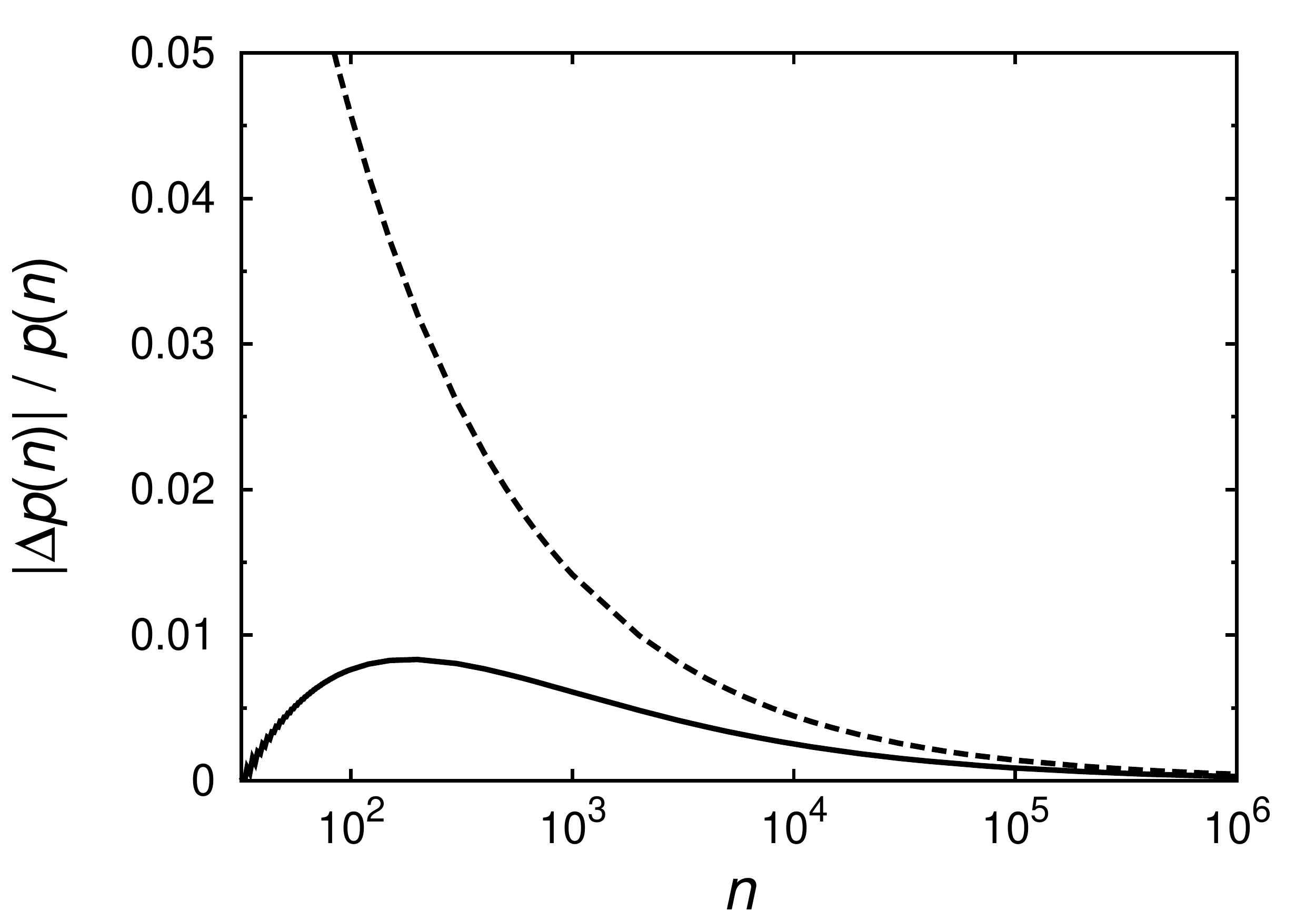}}
\caption{(From \cite{Prokhorov&Rovenchak:2012}.) Comparison of relative errors for the estimations of the number of integer partitions. Solid curve is result (\ref{p1d-corr}), dashed curve is the leading asymptotic provided by the Hardy--Ramanujan formula (\ref{eq-HR}).
}\label{fig:corr1D}
\end{figure}

For linear partitions this gives the following expression
\be\label{p1d-corr}
p_{\rm corr}(n) = \frac{1}{18\sqrt[4]{6}\, n^{3/4}}
e^{{28\over 27} \pi \sqrt{2n\over3}} 
K_{1/3}\left({1\over 27} \pi \sqrt{2n\over3}\right)
\end{align}
containing the first correction to the asymptotic result of Hardy and Ramanujan (\ref{eq-HR}) in a concise form and providing a better estimate ($< 1\%$ accuracy) even for small values of $n > 20$, see Fig.~\ref{fig:corr1D}.

In the two-dimensional case, the correction to the main asymptotic is easily evaluated in the same fashion yielding for plane partitions
\be\label{p2d-corr}
p^{\rm 2D}_{\rm corr}(n) = 
\frac{[2\zeta(3)]^{13/16}}{4\sqrt3\,\pi}\,n^{-13/36}
e^{\frac{25}{16}[2\zeta(3)]^{1/3}n^{2/3}+c}\:
K_{1/3}\left( \frac{[2\zeta(3)]^{1/3}\,n^{2/3}}{16} \right).
\end{align}

\begin{figure}[h]
\centerline{\includegraphics[scale=.75,angle=0]{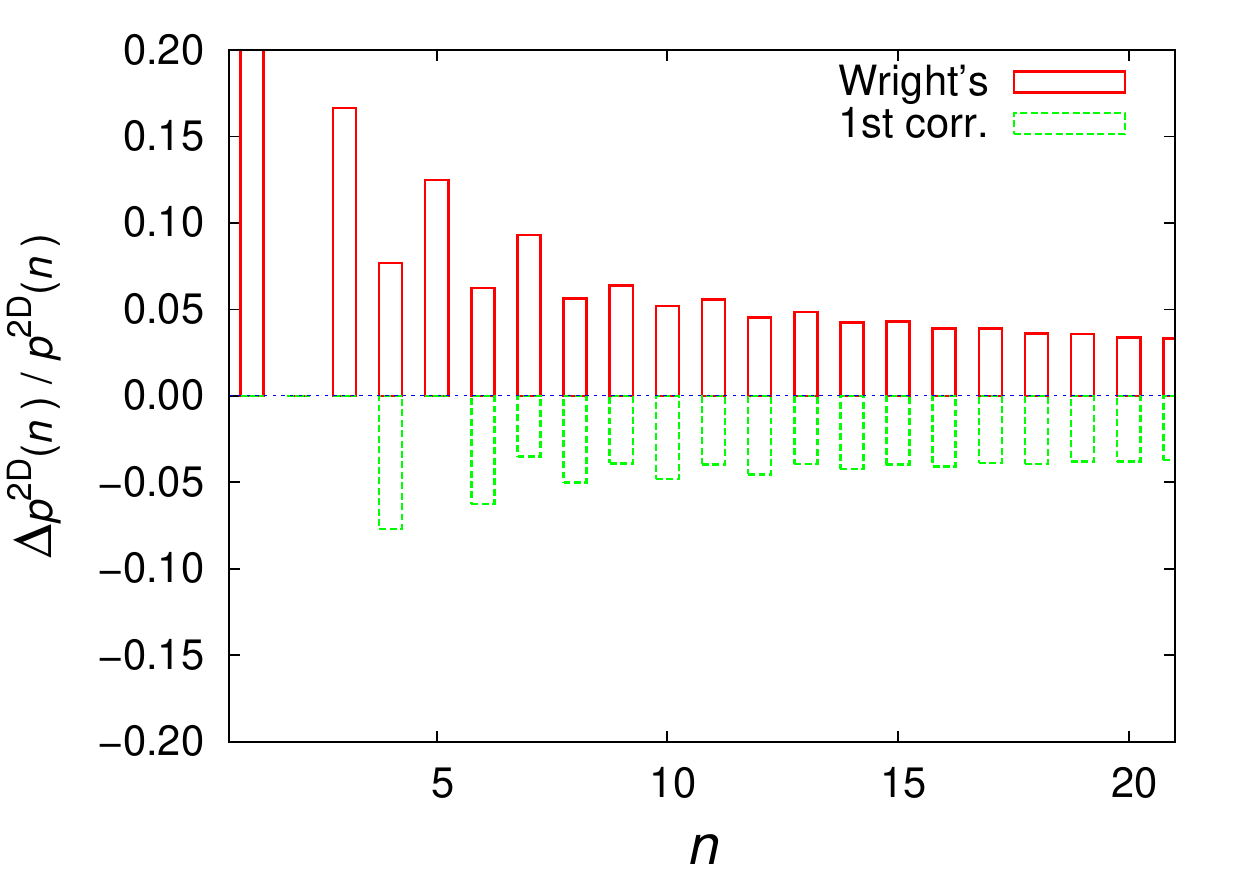}}
\caption{Comparison of relative errors for the estimations of the number of plane partitions. Solid (red) contour boxes with positive values correspond to the main asymptotic by Wright (\ref{p2d-Wright'}), 
dashed (green) ones with negative values represent the correction given by (\ref{p2d-corr}).
}\label{fig:corr2D}
\end{figure}

However, unlike the linear case, this correction provides a better estimate in comparison with the main asymptotic by Wright (\ref{p2d-Wright'}) only for small values of $n<17$. The relative error for $n=50$ is $+1.81\%$ for the main asymptotic and $-2.72\%$ for the correction given by (\ref{p2d-corr}), for $n=100$ the respective errors are $+1.13\%$ and	$-1.98\%$, for $n=1000$ they are $+0.24\%$ and $-0.54\%$, etc. The domain of small $n$ is illustrated in Fig.~\ref{fig:corr2D}.

\section{Restricted partitions}\label{sec:Restricted}
As it was mentioned in Sec.~\ref{sec:Phys}, integer partitions with a restricted number of parts are equivalent to a finite system of bosonic harmonic oscillators. The partition function $Z_N$ of such a system of $N$ particles obeys the following recurrence relation \cite{Borrmann&Franke:1993}:
\be\label{eq:ZN-recurr}
Z_N(x) = \frac{1}{N}\sum_{k=1}^N B_{k}(x)Z_{N-k}(x),\qquad 
Z_0(x)\equiv 1,
\end{align}
where $x={\rm e}^{-\beta\hbar\omega}$. In $D$ dimensions,
\be
B_k(x) = \frac{1}{(1-x^k)^D}.
\end{align}

\subsection{Linear partitions with a restricted number of parts}
A closed-form ex\-pres\-sion for $Z_N$ exists only in the one-dimensional case:
\be\label{eq:ZN1D}
Z^{\rm 1D}_N(x) = \prod_{k=1}^N \frac{1}{1-x^k}.
\end{align}

The partition function $Z(x)$ of an infinite system can be written as:
\be\label{eq:ZN=Zy}
Z_N(x) = Z(x) y_N(x),
\end{align}
where the function $y_N(x)$ has an obvious limiting behavior 
$\ds
\lim_{N\to\infty}y_N(x) = 1.
$
Thus, in the one-dimensional case,
\be
Z^{\rm 1D}(x) = \prod_{k=1}^N \frac{1}{1-x^k}\prod_{k=N+1}^\infty \frac{1}{1-x^k}
\end{align}
and consequently,
\be
y^{\rm 1D}_N(x) = \prod_{k=N+1}^\infty (1-x^k) = 
\exp\sum_{k=N+1}^\infty \ln\left(1-x^k\right).
\end{align}
Since the entropy of a finite system is $S_N = \beta E+\ln Z_N = \beta E+\ln Z +\ln y_N$, for the number of restricted partitions we immediately obtain from Eq.~(\ref{eq:GammaE}):
\be
p_N(n) = p(n) y_N({\rm e}^{-\beta_0}),
\end{align}

The finite-size correction is thus: 
\be\label{eq:yN-exact1}
y^{1\rm D}_N(x) = \exp\left(-\frac{x^{N}}{1-x}\right) \simeq  \exp\left(-\frac{x^{N}}{\beta_0}\right).
\end{align}
In the one-dimensional problem the stationary point is $\beta_0=\pi/\sqrt{6n}$.  Therefore, the leading correction in the number of restricted partitions reproduces the classical result (\ref{eq:EL-ln}) of Erd\H{o}s and Lehner \cite{Erdos&Lehner:1941}
\be
p_N(n) =  p(n) \exp\left\{-\frac{\sqrt{6n}}{\pi}\,{\rm e}^{-\pi N/\sqrt{6n}}\right\},
\end{align}
about the asymptotic behavior of the number of partitions of $n$ into at most $N$ parts.

\subsection{Plane partitions with a restricted number of parts}

The two-dimensional problem can be treated using the $Z$-transform techniques
to solve the recurrence relation (\ref{eq:ZN-recurr}) for the partition function.
This approach was suggested in  \cite{Rovenchak:2014TMF}.

The expression for the partition function of a system of $N$ isotropic 2D oscillators can be written as:
\be
\ln Z^{\rm 2D}_N(x) = -\sum_{k=1}^N k\ln(1-x^k),
\end{align}
At $N\to\infty$, this expression recovers MacMahon's generation function for plane partitions \cite{MacMahon:1896,Stanley:1971}:
\be
\sum_{n=0}^\infty p^{\rm 2D}(n) x^n = \prod_{n=1}^\infty \frac{1}{(1-x^n)^n}.
\end{align}
This yields for the correction $y^{\rm 2D}_N(x)$:
\be
y^{\rm 2D}_N(x) = \exp\left(-\frac{N x^N}{1-x}\right) \simeq \exp\left(-\frac{N x^N}{\beta_0}\right)
\end{align}
with the stationary point 
$\ds\beta_0 = [{2\zeta(3)}/{n}]^{1/3}$.

The following asymptotic behavior can thus be conjectured for the number of restricted plane partitions \cite{Rovenchak:2014TMF}:
\be\label{eq:MAIN}
p^{\rm 2D}_N(n) = p^{\rm 2D}(n) 
\exp\left\{-\frac{N n^{1/3}}{[2\zeta(3)]^{1/3}} {\rm e}^{-N [2\zeta(3) / n]^{1/3}}\right\}.
\end{align}
This formula gives the estimation of the number of plane partitions of $n$ into at most $N$ parts, where 
$0.75 n^{1/3} \ll N < n$
(this condition is due to limiting transitions in derivation procedures).

The above approach was tested for $n\leq20$ with $N/n=9/10$ yielding decreasing relative errors of $<5$--10\%. The conjectured expression (\ref{eq:MAIN}) yet awaits the numerical check in a wide range of the $n$ and $N$ values.

\subsection{Numerical results for restricted plane partitions at intermediate values of $\boldsymbol{N\sim n^{2/3}}$}

Numerical analysis of the recurrence relation (\ref{eq:ZN-recurr}) for $Z_N(x)$ in two dimensions was made for $n=100\div10\,000$. The value of the argument was $x=e^{-\beta_0}$ with the stationary point corresponding to the two-dimensional case as above, $\ds\beta_0 = [{2\zeta(3)}/{n}]^{1/3}$. Some computed values are presented in Table~\ref{tab:2}.

\begin{table}[h]
\caption{Solutions of the recurrence relation (\ref{eq:ZN-recurr}) for $\ln Z_N$ in the two-dimensional case}\label{tab:2}\medskip

\vspace*{-0.5cm}
\begin{footnotesize}
\begin{center}
\hspace*{-0.3cm}
\begin{tabular}{r|rrrrrrr}											
\hline\hline															
\backslashbox{\hspace*{-0.0cm}$N$}{\hspace*{-0.3cm}$n$} &	100\ \ \ \ &	200\ \ \ \ &	500\ \ \ \ &	1000\ \ \ &	2000\ \ \ \ &	5000\ \ \ \ &	10000\ \ \ \\
\hline																
10	&	13.50490	&	17.20871	&	22.46359	&	26.62624	&	30.90153	&	36.67638	&	41.11207	\\
20	&	17.11408	&	23.66709	&	33.53210	&	41.58209	&	49.96198	&	61.38032	&	70.19437	\\
30	&	18.09474	&	26.52429	&	40.27325	&	51.91522	&	64.22143	&	81.14853	&	94.28280	\\
50	&	18.30640	&	28.02642	&	47.03409	&	64.86995	&	84.45795	&	111.98343	&	133.58019	\\
100	&		&	28.13457	&	49.79461	&	75.97533	&	109.82935	&	161.18825	&	202.85769	\\
200	&		&		&	49.81234	&	76.99794	&	119.32409	&	203.58562	&	280.20762	\\
300	&		&		&		&	76.99798	&	119.40093	&	213.60298	&	316.40845	\\
500	&		&		&		&		&		&	214.27879	&	334.40026	\\
\hline																
$N\to n$	&	18.30884	&	28.13458	&	49.81234	&	76.99798	&	119.40094	&	214.27879	&	334.64773	\\
\hline\hline															
\end{tabular}													
\end{center}						
\end{footnotesize}

\end{table}
		
As one can notice, the values of $Z_N$ rapidly approach asymptotic results corresponding to the limit of $N\to n$. This significantly complicates the numerical analysis of the recurrence relation  (\ref{eq:ZN-recurr}) in the case discussed in the previous subsection.

Upon transferring these results to the problem of plane partitions, the following relation was found to describe the data quite satisfactorily
for intermediate values of $N\sim n^{2/3}$:
\be
\ln\frac{p_N^{\rm 2D}(n)}{p^{\rm 2D}(n)}=
-A n^{2/3} N^{-1/3} \exp\left\{-kN^{3/2}/n+b\,n^{1/3}\ln N\right\}.
\end{align}
Constants $A$, $b$, and $k$ were calculated as fitting parameters with the following values:
$A=1.075\pm0.008$, $b=0.0060\pm0.0002$, and $k=2.26\pm0.2$. Interestingly, the computed value of $k$ is close to $k=2\zeta(3)=2.40411\ldots$ making the respective factor $k/n=\beta_0^3$.

Therefore, the second asymptotic relation can be conjectured for the number of restricted plane partitions in the following form:
\be
\ln\frac{p_N^{\rm 2D}(n)}{p^{\rm 2D}(n)}\sim
-\left(\frac{n^2}{N}\right)^{1/3} 
\exp\left\{-\frac{2\zeta(3)N^{3/2}}{n}\right\}
\end{align}
for the number of parts $N\sim n^{2/3}$.

\section{Final remarks}

In summary, an overview of some problems in the theory of integer partitions was given and relevant results obtained within the statistical mechanics ap\-proach were briefly discussed. Known asymptotic expressions for unrestricted linear and plane partitions were reproduced and the possibility to obtain the correction to the leading asymptotic was demonstrated.

The asymptotic expression for the number of restricted linear partitions into at most $N$ parts was correctly reproduced as well. Asymptotic formulas for the number of plane partitions into at most $N$ parts were conjectured. The asymptotic behavior of restricted plane partitions with $N\sim n^{2/3}$ was estimated numerically. The conjectured formulas await testing for large $n$ and $N$.

Possible generalization for higher-order partitions can be derived within the discussed approaches, with some preliminary analysis presented in \cite{Rovenchak:2010CMST}. Application to physical problems, especially to study quantum ensembles of particles obeying different types of statistics, is seen as one of the principal directions relating integer partitions and statistical mechanics, which require further exploration.

\subsection*{Acknowledgments}
I am grateful to the organizers of the conference ``Arithmetic Me\-thods in Mathematical Physics and Biology'' for the opportunity to present these results at such a highly interdisciplinary meeting.



\begin{thebibliography}{00}








\bibitem{Andrews:1976}G. E. Andrews, \emph{The Theory of Partitions}, Addison-Wesley, Reading, Mass., 1976.

\bibitem{Apostol:1999}T. M. Apostol, 
\emph{An elementary view of Euler's summation formula},
Am. Math. Mon. {106} (1999), 409--418.

\bibitem{Auluck&Kothari:1946}F. C. Auluck and D. S. Kothari,
\emph{Statistical mechanics and the partitions of numbers},
Proc. Camb. Phil. Soc. {42} (1946), 272--277.

\bibitem{Bogoliubov:2007}N. M. Bogoliubov, 
\emph{Enumeration of plane partitions and the algebraic Bethe anzatz}, Theor. Math. Phys. {150} (2007) 165--174.

\bibitem{Bohr&Kalckar:1937}N. Bohr and F. Kalckar, 
\emph{On the transmutation of atomic nuclei by impact of material particles. I. General theoretical remarks}, Kgl. Danske Vid. Selskab. Math. Phys. Medd. {14}, No.\,10 (1937), 1--40.

\bibitem{Borrmann&Franke:1993}P. Borrmann and G. Franke,
\emph{Recursion formulas for quantum statistical partition functions},
J.~Chem. Phys. {98} (1993), 2484--2485.

\bibitem{Chatterjee&Diaconis:2014}S. Chatterjee and P. Diaconis,
\emph{Fluctuations of the Bose--Einstein condensate},
J.~Phys. A: Math. Theor. {47} (2014), 085201 (23~p.).

\bibitem{Comtet_etal:2007}A. Comtet, P. Leboeuf, and S. N. Majumdar, 
\emph{Level density of a Bose gas and extreme value statistics},
Phys. Rev. Lett. {98} (2007), 070404 (4~p.).

\bibitem{Debnath:1987}L.~Debnath,
\emph{Srinivasa Ramanujan (1887--1920) and the theory of
partitions of numbers and statistical mechanics. A centennial tribute},
Internat. J. Math. Math. Sci. {10}, (1987) 625--640.

\bibitem{Dutta:1953}M.~Dutta,
\emph{An essential statistical approach to thermodynamic problem II}, Proc. Natl. Inst. Scl. India A {19} (1953), 109--126.

\bibitem{Dutta:1956}M.~Dutta,
\emph{On new partition of numbers},
Rend. Semin. Mat. Univ. Padova {25} (1956), 138--143.

\bibitem{Erdos&Lehner:1941}P. Erd\"os and J. Lehner, 
\emph{The distribution of the number of summands in the partitions of a positive integer},
Duke Math.~J. {8} (1941), 335--345.

\bibitem{Euler:1751}L. Euler, \emph{Observationes analyticae variae de combinationibus}, 
Commentarii academiae scientiarum Petropolitanae {13} (1751), 64--93.

\bibitem{Euler:1753}L. Eulero, \emph{De partitione numerorum},
Novi Commentarii Academiae scientiarum Petropolitanae {3} (1753), 125--169.

\bibitem{Grossmann&Holthaus:1996}S. Grossmann and M. Holthaus,
\emph{Microcanonical fluctuations of a Bose system's ground state occupation number}, Phys. Rev. E {54} (1996), 3495--3498.

\bibitem{Grossmann&Holthaus:1997}S. Grossmann and M. Holthaus,
\emph{Fluctuations of the particle number in a trapped Bose--Einstein condensate}, Phys. Rev. Lett. {79} (1997), 3557--3560.

\bibitem{Grossmann&Holthaus:1999}S. Grossmann and M. Holthaus,
\emph{From number theory to statistical mechanics: Bose–Einstein condensation in isolated traps}, Chaos Solitons Fractals {10} (1999), 795--804.

\bibitem{Abramowitz&Stegun:1972}
\emph{Handbook of Mathematical Functions.} Tenth printing,
M.~Abramowitz and I.~A.~Stegun (eds.),
National Bureau of Standards, 1972.

\bibitem{Hardy&Ramanujan:1917}G.-H. Hardy and S. Ramanujan,
\emph{Une formule asymptotique pour le nombre des partitions de $n$}, 
C.~R.~Hebd. S\'eances Acad. Sci. {164} (1917), 35--38.

\bibitem{Hardy&Ramanujan:1918}G. H. Hardy and S. Ramanujan,
\emph{Asymptotic formul\ae\ in combinatory analysis}, 
Proc. London Math. Soc. {17} (1918), 75--115.

\bibitem{Leibniz:1674}G. W. Leibniz,
\emph{Specimen de divulsionibus aequationum ad problemata indefinita in numeris rationalibus solvenda. 2. September 1674},
in: G. W. Leibniz, \emph{S\"amtliche Schriften und Briefe. Siebente Reihe: Mathematische Schriften, Bd.~1: 1672--1676. Geometrie -- Zahlentheorie -- Algebra (1.~Teil)}, Akademie-Verlag, Berlin, 1990, 740--755.

\bibitem{MacMahon:1896}P.~A.~MacMahon, 
\emph{Memoir on the theory of the partition of numbers. Part I},
Phil. Trans. R. Soc. London A, 187 (1896), 619--673.

\bibitem{MacMahon:1915}P.~A.~MacMahon, 
\emph{Combinatory Analysis}, 2 vols, Cambridge University Press, 1915--1916.

\bibitem{Nanda:1951}V. S. Nanda,
\emph{"Partition theory and thermodynamics of multi-dimensional oscillator assemblies},
Proc. Camb. Phil. Soc. {47} (1951), 591--601.

\bibitem{Prokhorov&Rovenchak:2012}D. Prokhorov and A. Rovenchak, 
\emph{Asymptotic formulas for integer partitions within the approach of microcanonical ensemble},
Condens. Matter Phys. {15} (2012), 33001 (9~p.). 

\bibitem{Rademacher:1938}H.~Rademacher,
\emph{On the partition function $p(n)$},
Proc. London Math. Soc. (Ser.~2) {43} (1938), 241--254.

\bibitem{Roccia&Leboeuf:2010}J. Roccia, P. Leboeuf,
\emph{Level density of a Fermi gas and integer partitions: A Gumbel-like finite-size correction}, Phys. Rev. C {81} (2010), 044301 (5~p.).

\bibitem{Rovenchak:2009FNT}A. Rovenchak, 
\emph{The relation between fractional statistics and finite bosonic systems in one-dimensional case},
Low Temp. Phys. {35} (2009), 400--403; 
Fiz. Nizk. Temp. {35} (2009), 510--513.

\bibitem{Rovenchak:2010CMST}A. Rovenchak, 
\emph{Partition function formalism in the problem of multidimensional integer partitions},
Computat. Meth. Sci. Technol. (Pozna\'n) {16} (2010), 187--190. 

\bibitem{Rovenchak:2014TMF}A. Rovenchak, 
\emph{Enumeration of plane partitions with a restricted number of parts},
Theor. Math. Phys. {181} (2014), 1428--1434. 

\bibitem{Stanley:1971}R. P. Stanley,
\emph{Theory and application of plane partitions. Parts 1 and 2}, Stud. Appl. Math. {50} (1971), 167--188 and 259--279.

\bibitem{Temperley:1949}H. N. V. Temperley, 
\emph{Statistical mechanics and the partition of numbers. I.~The transition of liquid helium}, 
Proc. R. Soc. London A {199} (1949), 361--375.

\bibitem{Tran_etal:2004}M. N. Tran, M. V. N. Murthy, and R. J. Bhaduri,
\emph{On the quantum density of states and partitioning an integer}, Ann. Phys. {311} (2004), 204--219.

\bibitem{VanLier&Uhlenbeck:1937}C. Van Lier and G. E. Uhlenbeck,
\emph{On the statistical calculation of the density of the energy levels of the nuclei}, Physica {4} (1937), 531--542.  

\bibitem{Wright:1931}E. M. Wright,
\emph{Asymptotic partition formulae: I. Plane partitions},
Quart. J. Math. Oxford Ser. {2} (1931), 177--189.

\end{thebibliography}
\end{document}